\documentclass[aps,prd,twocolumn,nofootinbib,floatfix,showpacs]{revtex4-1}
\usepackage[latin1]{inputenc}  
\usepackage{bbm,slashed}
\usepackage{mathrsfs} 
\usepackage{graphicx,epsfig}
\usepackage{amsmath,amsfonts,amssymb}
\usepackage{booktabs} 

\usepackage{color}

\def\Z{\mathbbm{ Z}}

\newcommand{\Q}{\mathcal{Q}}

\newcommand{\psib}{\bar{\psi}}

\newcommand{\Dbar}{\bar{\D}} 
 
\newcommand{\N}{\mathcal{N}} 
\newcommand{\D}{\mathcal{D}}

\newcommand{\pa}{\partial}

\newcommand{\ha}{\frac{1}{2}}

\newcommand{\chib}{\bar\chi}

\newcommand{\w}{\mathfrak w}

\newcommand{\Li}{\mathrm {Li}}
\newcommand{\T}{\tilde{T}}

\graphicspath{{plots/}}

 \DeclareMathAlphabet{\boldmathe}{T1}{cmr}{bx}{it}

\newcommand{\id}{\mathbbm{1}}

\newcommand{\lam}{\lambda}

\def\Str{\mathrm{Str}}

\newcommand{\PfadD}{\mathcal{D}}
\newcommand{\ew}[1]{\langle{#1}\rangle}	
\newcommand{\functder}[2]{\frac{\delta #2}{\delta #1}}
\newcommand{\functderder}[3]{\frac{\delta^2 #3}{\delta #1\delta #2}}
\graphicspath{{plots/}}

\begin{document}
\title{${\cal N}=1$ Wess Zumino Model in $d=3$ at zero and finite
temperature}

\author{Franziska Synatschke, Jens Braun and Andreas
  Wipf}
\affiliation{
Theoretisch-Physikalisches Institut, Friedrich-Schiller-Universit{\"a}t Jena, 
Max-Wien-Platz~1, D-07743~Jena, Germany}

\begin{abstract}
Supersymmetric renormalization group (RG) flow equations for the effective 
superpotential of the three-dimensional Wess-Zumino model are derived  
at zero and non-zero temperature. This model with fermions and bosons 
interacting via a Yukawa term possesses a supersymmetric analogue of the 
Wilson-Fisher fixed-point. At zero temperature we determine the phase-transition 
line in coupling-constant space separating the supersymmetric from the 
nonsupersymmetric phase. At finite temperature
we encounter dimensional reduction from $3$ to $2$ dimensions 
in the infrared regime. We determine the finite-temperature phase diagram 
for the restoration of the global $\Z_2$-symmetry and
show that for temperatures above the $\Z_2$ phase-transition the pressure obeys 
the Stefan-Boltzmann law of a gas of massless bosons in $2+1$ dimensions.

\end{abstract}
\pacs{05.10.Cc,12.60.Jv,11.30.Qc,11.10.Wx}
\maketitle  

\section{Introduction}
In this paper we investigate the \emph{three-dimensional} \mbox{$\mathcal N=1$}
Wess-Zumino model with general superpotential and explore the model beyond the realm of perturbative 
expansions. This quantum field theory describes Majorana fermions and uncharged 
bosons in interaction, with their spatial motion restricted to a two-dimensional layer. 
The selfcoupling of the bosons and the Yukawa coupling between fermions and bosons 
are such that the theory possesses one supersymmetry.  There exists a class of
superpotentials for which the three-dimensional models 
are perturbatively renormalizable, in contrast to the four-dimensional models.
For superpotentials of the form $W(\phi)\sim \lam\phi^{2n+1}$ there exists both a 
supersymmetric and a nonsupersymmetric phase. 
In this paper we shall calculate the phase-transition curve
separating the supersymmetric from the nonsupersymmetric phase. Besides
supersymmetry the action is invariant under $\phi\to-\phi$, and at zero
temperature the breaking of this global ${\mathbb Z}_2$ symmetry is 
intimately linked to the breaking of supersymmetry. We shall see that there exists a finite 
phase-transition temperature at which ${\mathbb Z}_2$ symmetry is restored, 
independent of our choice for the couplings at the cutoff-scale. 
Similarly as for other two-dimensional systems, e.g. surface science, 
heterostructures or electron gases, the physics in two space-dimensions is rather 
different from that in three space-dimensions.

We employ the functional renormalization group (RG) to calculate the
phase structure at zero and finite temperature, the scaling behavior of 
the mass with the RG-scale, the wave-function renormalization, critical exponents, the effective
potential and the temperature dependence of the pressure. The
method has previously been applied to a wide range of nonperturbative problems 
such as critical phenomena, fermionic systems, gauge theories and quantum gravity, 
see  \cite{Litim:1998nf,Aoki:2000wm,Berges:2000ew,Polonyi:2001se,Pawlowski:2005xe,Gies:2006wv,Sonoda:2007av,Delamotte:2007pf}
for reviews. A number of conceptual studies of supersymmetric theories has
already been performed with the functional RG. The delicate point here is, of
course, the construction and use of a manifestly supersymmetry-preserving
regulator.  For the four-dimensional Wess-Zumino model such
a regulator has been presented in \cite{Vian:1998kv,Bonini:1998ec}. 
Recently, general theories of a scalar superfield including the Wess-Zumino 
model have been investigated with
a Polchinski-type RG equation in \cite{Rosten:2008ih}, yielding a new approach
to supersymmetric nonrenormalization theorems. A Wilsonian effective action
for the Wess-Zumino model by perturbatively iterating the functional RG has
been constructed in \cite{Sonoda:2008dz}.

The present study  builds on our earlier results on two-dimensional
supersymmetric field theories at zero temperature \cite{Synatschke:2009nm,Gies:2009az} 
as well as on supersymmetric quantum mechanics, where we have constructed a  
manifestly supersymmetric functional RG flow, see \cite{Synatschke:2008pv}.
The two-dimensional models possess an infinite series of fixed points
described by two-dimensional super-conformal theories. On the contrary, 
supersymmetric models in three dimensions possess just one fixed point, 
similarly as three-dimensional $O(N)$ models, see e. g.~\cite{Berges:2000ew,Tetradis:1993ts,Litim:2002cf,Bervillier:2007rc}.

In the present work we first sketch the derivation of the manifestly supersymmetric
flow equations in Sect.~\ref{sec:FEZT}. Since there exist no 
Majorana fermions in three-dimensional Euclidean spacetime we derive the flow equations in Minkowski 
spacetime and continue the result to imaginary time. We investigate the flow of the
superpotential in Sect.~\ref{subsec:LPA}, study the fixed-point structure in 
detail, and identify the supersymmetric analogue of the Wilson-Fisher fixed point 
of three-dimensional bosonic $O(N)$ models with one unstable direction. Taking
into account a nonzero anomalous dimension in Sect. \ref{sec:NTLO} yields a scaling relation
between the critical exponent of the unstable direction and the anomalous
dimension. In addition we determine the zero-temperature phase diagram 
for spontaneous breaking of supersymmetry.

The second part of this paper is devoted to the behavior of the model at
finite temperature. The fate of supersymmetry at finite temperature
has been discussed extensively in the literature. For example, in previous works
the KMS condition has been implemented directly in thermal superspace
\cite{Derendinger:1998zj}. In \cite{Das:1978rx,Girardello:1980vv} the supersymmetry 
breaking has been studied on the level of thermal Green functions.
The breaking of supersymmetry by finite temperature corrections, for
example the one-loop corrections to fermionic and bosonic masses, has
been determined in the real-time formulation in \cite{Midorikawa:1984fi}.
The inevitable breaking of supersymmetry at finite temperature has sometimes
been called \emph{spontaneous collapse} of supersymmetry \cite{Buchholz:1997mf}.

In Sect. \ref{sec:FEAFT} we derive the RG flow equations at finite temperature.
In addition to the momentum integrals we are confronted with
sums over Matsubara frequencies. For the three-dimensional Wess-Zumino model and for a 
particular regulator the thermal sums can be calculated analytically. Related sums have 
been discussed in earlier works on finite-temperature renormalization group flow equations, 
for example in \cite{Litim:2000ci,Litim:2001up,Braun:2006jd,Litim:2006ag,Blaizot:2006rj,Floerchinger:2008qc,Braun:2009si,Diehl:2009ma}.
We observe that the Wess-Zumino model in three dimensions at finite temperature
in the $\Z_2$ symmetric phase behaves similarly to a gas of massless
bosons. In particular we show in Sect.~\ref{sec:pressure} that it obeys the Stefan-Boltzmann law in three dimensions. 
For high temperatures the fermions do not contribute to the flow equations since
they do not have a thermal zero-mode. On the other hand we observe dimensional reduction
in the bosonic part of the model due to the presence of a thermal zero-mode. We 
show in Sect. \ref{sec:highT} how this is manifested in our RG framework.
In a similar way dimensional reduction has been observed in $O(N)$-models at finite temperature 
in \cite{Tetradis:1992xd,Bohr:2000gp}. Finally we compute the phase diagram  for the restoration 
of the global $\mathbb Z_2$ symmetry at finite temperature in Sect.~\ref{sec:PTPD}.

\section{The $\N=1$ Wess-Zumino model in three dimensions at $T=0$}
There are many works on the supersymmetric Wess-Zumino models in
both four and two space-time dimensions. Actually the two-dimensional
model with $\N=2$ supersymmetries is just the toroidal compactification
of the four-dimensional $\N=1$ model. The three-dimensional
model with $\N=1$ supersymmetry, on the other hand, cannot be obtained 
by dimensional reduction of a local field theory in four dimensions. Thus
it may be useful to recall the construction of the three-dimensional model
starting from the real superfield 
\begin{equation} 
  \Phi(x,\alpha)=\phi(x)+\bar\alpha\psi(x)
  	+\frac12\bar\alpha\alpha F(x)
  	\label{eq:model1}  
\end{equation}
with real (pseudo)scalar fields $\phi,F$ and Majorana spinorfield $\psi$.  The supersymmetry 
variations are generated by the supercharge 
\begin{equation}
  \delta_\beta \Phi=i\bar\beta \Q\Phi,\quad
  \Q=-i\frac{\partial}{\partial\bar\alpha}-(\gamma^\mu\alpha)\partial_\mu\,.
  \label{eq:model3}
\end{equation}
We use the metric tensor $(\eta_{\mu\nu})=\hbox{diag}(1,-1-1)$ to lower
Lorentz indices. With the aid of the symmetry relations for Majorana spinors
$
  \psib\chi=\chib\psi,\;  \psib\gamma^\mu\chi=-\chib\gamma^\mu\psi
$
and the particular Fierz identity $\alpha\bar\alpha=-\bar\alpha\alpha\,\id/2$
the transformation laws for the component fields follow from Eq.~\eqref{eq:model3}:
\begin{equation}
  \delta\phi=\bar\beta\psi,\quad
  \delta\psi=(F+i\slashed\partial \phi)\beta,\quad
  \delta F=i\bar\beta\slashed\partial\psi\,.\label{eq:model9}
\end{equation}
The anticommutator of two supercharges yields
$\left\{\Q_\alpha,\bar \Q^\beta\right\}=2{(\gamma^\mu)_\alpha}^\beta
  \partial_\mu$. The supercovariant derivatives are 
\begin{equation}
  \D=\frac{\partial}{\partial\bar\alpha}+i(\gamma^\mu\alpha)\partial_\mu,\quad
  \text{and}\quad
   \Dbar=-\frac{\partial}{\partial\alpha}-i(\bar\alpha\gamma^\mu)\partial_\mu.
  \label{eq:model11}
\end{equation}
Up to a sign they obey the same anticommutation relation as the supercharges
\begin{equation}
  \{\D_\alpha,\Dbar^\beta\}=-2{(\gamma)_\alpha}^\beta
  \partial_\mu\,.\label{eq:model13}
\end{equation}
As kinetic term we choose the $D$ term of 
$\Dbar \Phi\D\Phi=2\bar\alpha\alpha\mathcal{L}_{\rm kin}+\dots$ 
which reads
\begin{equation}
  \mathcal L_{\rm kin}=\frac12\partial_\mu\phi\partial^\mu\phi-
  	\frac i2\psib\slashed\partial\psi
  	+\frac12 F^2.
\end{equation}
The interaction term is the $D$ term of  $2W(\Phi)=\bar\alpha\alpha \mathcal{L}_{\rm int}+\dots$ 
and contains a Yukawa term,
\begin{equation}
	\mathcal L_{\rm int}=FW'(\phi)-\frac12W''(\phi)\psib\psi.
\end{equation}
The complete off-shell Lagrange density  $\mathcal L_{\rm off}=\mathcal L_{\rm kin}
+\mathcal L_{\rm int}$ takes then the simple form
\begin{equation}
  \mathcal L_{\rm off}
  =\frac12\partial_\mu\phi\partial^\mu\phi-
  	\frac i2\psib\slashed\partial\psi
  	+\frac12 F^2+FW'(\phi)-\frac12W''(\phi)\psib\psi.
\end{equation}
Eliminating the auxiliary field via its equation of motion
$F=-W'(\phi)$,
we end up with the on-shell density
\begin{align}
\mathcal L_{\rm on}&=\frac12\partial_\mu\phi\partial^\mu\phi-\frac
i2\psib\slashed\partial\psi-\frac12W'^2(\phi)-\frac12W''(\phi)\psib\psi.
\end{align}
From this expression we read off that $W'^2(\phi)$ acts as a self-interaction 
potential for the scalar fields. For a polynomial superpotential $W(\phi)$ in which 
the power of the leading term is even, $W(\phi)=c\phi^{2n}+\mathcal O(\phi^{2n})$, 
we do not observe supersymmetry breaking in our present non-perturbative renormalization
group study\footnote{ In a two-loop calculation a ground state with
broken supersymmetry has been found in Ref.~\cite{Lehum:2008vn}. Since we neglect
higher $F$-terms in our non-perturbative study it is not possible to check whether the findings
of this perturbative analysis of the Wess-Zumino model hold when higher-order corrections 
are taken into account.}. On the other hand spontaneous supersymmetry breaking is definitely possible for 
a superpotential in which the power of the leading term is odd. 
In the explicit calculations we shall use a Majorana representation for the $\gamma$-matrices,
$\gamma^0=\sigma_2,\,
\gamma^1=i\sigma_3$ and $\gamma^2=i\sigma_1.$

\section{Flow equation at zero temperature}\label{sec:FEZT}
To find a manifestly \emph{supersymmetric flow equation} in the
off-shell formulation we extend our earlier results on the one- and two-dimensional
Wess-Zumino models \cite{Synatschke:2009nm,Synatschke:2008pv} to
three dimensions.  Since there are no Majorana fermions in 
three-dimensional Euclidean space we begin with a Minkowski spacetime
formulation of the Wetterich equation~\cite{Wetterich:1992yh,Berges:2008sr}:  
\begin{align}
	\partial_{t}\Gamma_k=\frac
	i2\Str\left[\left(\Gamma_k^{(2)}+R_k\right)^{-1}\partial_{t} R_k\right],
	\quad t=\ln k^2\,, \label{eq:flow1}
\end{align}
where the scale-dependent effective action $\Gamma_k$ interpolates 
between the microscopic (classical) action $\Gamma_{k=\Lambda}=S$ 
and the full quantum effective action $\Gamma_{k=0}=\Gamma$. 
The second functional derivative in Eq. \eqref{eq:flow1} is defined as 
\[
\left(\Gamma_k^{(2)}\right)_{ab}=\frac{\overrightarrow{\delta}}{\delta\Psi_a}
\Gamma_k\frac {\overleftarrow{\delta}}{\delta\Psi_b}\,,
\] 
where the indices $a,b$ denote the field components, internal and Lorentz 
indices, as well as space-time or momentum coordinates, i.e., $\Psi^{\text T}=(\phi,F,\psi,\bar\psi)$ 
is a vector of fields, \emph{not}  to be confused with a superfield.
The cutoff function $R_k$ provides an infrared (IR) cut-off for all fields and specifies
the Wilsonian momentum-shell integrations such that the flow of $\Gamma_k$ is 
dominated by modes $p^2 \simeq k^2$. For a derivation\footnote{In the following we neglect
an additional term to the flow equation arising from the normalization of the Gau\ss ian
measure in the partition function. Including such an additional term yields a field-independent
constant to $W'(\phi)$. We stress that only non-universal quantities such as the
critical temperature for the $\Z_2$ phase transition are affected by such a constant.
However, our analysis of the critical dynamics at the phase boundary at zero and finite 
temperature is not affected.} of the RG
flow equation in Minkowski space-time~\eqref{eq:flow1} we refer to
App.~\ref{sec:MinkowskiERG}.

To construct a supersymmetric flow we use as regulator an invariant
$D$ term in superspace. Since such a term should be quadratic 
in the fields\footnote{A Regulator term quadratic 
in the fluctuation fields ensures that we eventually obtain a non-perturbative 
RG equation with one-loop structure.}, see App.~\ref{sec:MinkowskiERG} for details, it
is the $D$ term of a superfield $\Phi K\Phi$ with $K$ being a 
function of $\Dbar\D$. 
Using the anticommutation relation \eqref{eq:model13}, powers of $\Dbar\D$ can
always be decomposed into 
\begin{equation}
  \Big(\frac12\Dbar\D\Big)^{2n}=(-\Box)^n\,,
\end{equation}
such that an invariant and quadratic regulator action is the superspace
integral of 
\begin{equation}
\Phi K(\Dbar\D)\Phi=
\Phi \left(
r_1(-\Box)-r_2(-\Box)\frac{\Dbar \D}{2}\right)\Phi.
\end{equation}
Expressed in component fields, we find
\begin{equation} 
\Delta S_k=\ha\int(\phi,F) R^{\rm B}_k\, {\phi \choose F}
+\ha\int \psib R^{\rm F}_k\psi.\label{eq:lpa11}
\end{equation}
In momentum space, $i\pa_\mu$ is replaced by $p_\mu$ and the operators take the
explicit forms
\begin{equation}
R_k=\begin{pmatrix}
    R_k^{\rm B}&0\\0&R_k^{\rm F}
    \end{pmatrix}
\end{equation}
with
\begin{equation}
R_k^{\rm B}=\begin{pmatrix}
      p^2r_2&r_1\\r_1&r_2
      \end{pmatrix},\quad\text{and}\quad
R_k^{\rm F}=-r_1-r_2\slashed{p}\,,
\end{equation}
where $r_1=r_1(p^2/k^2)$ and $r_2=r_2(p^2/k^2)$. Note that the requirement that the 
RG flow preserves supersymmetry relates the regulators $R^{\rm B}_k$ and $R^{\rm F}_k$ in 
the bosonic and fermionic subsectors.

\subsection{Local Potential Approximation}\label{subsec:LPA}

We employ the following ansatz for the \emph{supersymmetric effective action} 
for  our study of the three-dimensional Wess-Zumino model:
\begin{equation}
  \begin{split}
 \Gamma_k
  =\int d^3x\left[\frac12\partial_\mu\phi\partial^\mu\phi-
  	\frac i2\psib\slashed\partial\psi
  	+\frac12 F^2\right.\\\left.+FW_k'(\phi)-\frac12W_k''(\phi)\psib\psi\right].
  	\label{eq:lpa1}
  	\end{split}
\end{equation}
In the following we work in the so-called local potential approximation (LPA) where 
the expectation values of the fields are taken to be constant over the entire volume. 
As it has been found in studies of scalar $O(N)$ models and low-energy QCD models, 
the LPA captures already most qualitative and quantitative features associated
with critical dynamics at zero and finite temperature provided the anomalous dimensions
are small, see e.~g. Refs~\cite{Tetradis:1993ts,Berges:2000ew,Bohr:2000gp,Litim:2002cf,Bervillier:2007rc,Braun:2009si}.
For the time being we restrict our study to LPA. In Sect.~\ref{sec:NTLO} we shall then discuss 
the running of the wave-function renormalization. 

In order to obtain a flow equation for the superpotential,
we project Eq. \eqref{eq:flow1} onto the terms linear in the auxiliary field 
$F$ and integrate the resulting $W'_k$ with respect to $\phi$. Performing a 
Wick rotation of the zeroth component of the momentum, i.\,e. $p_0^M\to
ip_0^E$, we find the flow equation
 \begin{equation}
    \partial_k W_k(\phi)=\frac12\int\frac{d^3 p}{(2\pi)^3}
    	\frac{\partial_k r_1(1+r_2)-\partial_k r_2 (W''_k(\phi)-r_1)}
    		{p^2(r_2+1)^2+(W''_k(\phi)+r_1)^2}\,.\label{eq:lpa3}
 \end{equation}
 In the following we choose the simple regulator functions
\begin{equation}
r_1=0,\quad r_2=\left({\frac{k}{|p|}}-1\right)\theta(k^2-p^2),
\label{eq:lpa4}
\end{equation}
 for which the momentum integration in \eqref{eq:lpa3}
 can be performed analytically. In the present work we do not aim at a
study the regulator dependence. However, the regulator dependence of functional RG 
flows, in particular with respect to critical phenomena,
has been investigated in great detail and it has been shown that optimized regulator
functions minimizing the trucational error can be constructed, see e.~g. 
Refs.~\cite{Litim:2000ci,Litim:2001up,Litim:2001hk,Litim:2002cf,Pawlowski:2005xe,Litim:2006ag}. 

Contrary to the model in two dimensions~\cite{Synatschke:2009nm}, the regulator 
function~\eqref{eq:lpa4} regularizes the flow even if we allow for running wave 
function renormalizations. For the superpotential $W_k$ we then obtain the simple flow 
equation
\begin{equation}
  \partial_kW_k=-\frac{k^2}{8\pi^2}\frac{W''_k(\phi)}{k^2+W''_k(\phi)^2}.
  	\label{eq:lpa5}
\end{equation}

As we are interested in the bosonic potential $V(\phi)= W'^2(\phi)/2$ we will
consider mostly the flow equation for $W'_k(\phi)$ which reads
\begin{equation}
 	\partial_kW_k'=-\frac{k^2 W_k^{(3)}(\phi ) \left(k^2-W_k''(\phi )^2\right)}{8
 	\pi ^2 \left(k^2+W_k''(\phi )^2\right)^2}. \label{eq:lpa7}
\end{equation}
\begin{figure}
\includegraphics[width=\columnwidth]{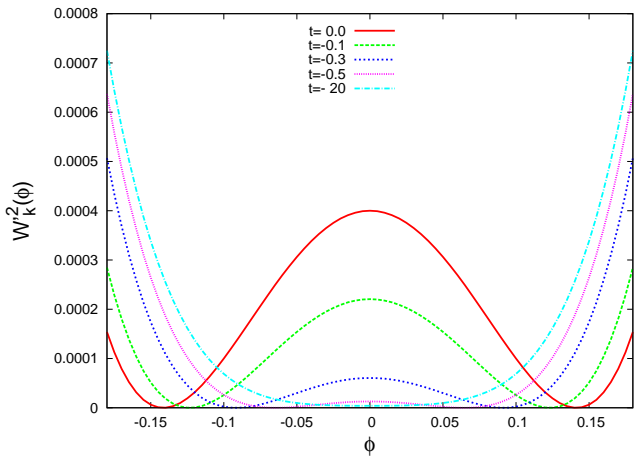}
\caption{RG flow of $W_k '^2(\phi)$ with the initial conditions $\lambda_\Lambda/\Lambda=1$, 
$a^2_\Lambda/\Lambda^{0.5}=0.02$. Note that $t$ is related to $k$ via $t=\ln k^2$, see also 
Eq.~\eqref{eq:flow1}.
\label{fig:FlowOfPotential}}
\end{figure}
Fig.~\ref{fig:FlowOfPotential} shows the flow of
$W'^2_k(\phi)$ for a quadratic superpotential at 
the cutoff scale, $W'_\Lambda=\lambda_\Lambda(\phi^2-a^2_\Lambda)$, 
and with initial conditions $\lambda_\Lambda/\Lambda=1$,
$a^2_\Lambda/\Lambda^{0.5}=0.02$. With these initial conditions
the RG flow starts in the regime with broken $\mathbb Z_2$ symmetry and for $k\to 0$ ends up in the regime with restored 
$\mathbb Z_2$ symmetry. We observe that the potential $V_k$ becomes 
flat  at the origin as $k$ is lowered to the infrared. 
In addition the function  $W''_{k\to0}$ is regular for all values of the field,
in contrast to the situation in  two dimensions.

In order to study the fixed-point structure we introduce 
dimensionless quantities
\begin{equation}
  \begin{split}
  \varphi=k^{-\frac12}\phi,\quad w_k(\varphi)=k^{-2}W_k(\phi),\\
  	w_k'(\varphi)=k^{-\frac32}W_k'(\phi),\quad \ldots\,.
  	\label{eq:fixPoints1}
  \end{split}	
\end{equation}
The dimensionless flow equation then reads
\begin{align}
	\partial_t w_k+2w_k=
		-\frac{w''_k}{8\pi^2(1+w''^2_k)}+\frac{\varphi w'_k}{2}\,,
\label{eq:fixPoints3}
\end{align}
and its fixed points are characterized by $\partial_tw_\ast=0$. The
flow equations in two and three dimensions have almost identical forms.
In three dimensions, however, there appears the additional 
term $\propto\varphi w'_k(\varphi)$, since the field $\phi$ itself is a dimensionful 
quantity. 

We observe a further  peculiarity of supersymmetric Wess-Zumino
models: Only the second derivative of the  superpotential 
enters the fixed-point equation following from $\pa_tw_\ast=0$,
see~\cite{Synatschke:2009nm}. It follows that the couplings of the terms $\phi^0$ and $\phi^1$ do  not enter
the fixed-point equation but evolve independently.
As we shall see below, this has some interesting consequences 
which distinguish the supersymmetric Wess-Zumino model from
purely bosonic theories, for example $O(N)$ models
in three dimensions, see e.\,g.
~\cite{Tetradis:1993ts,Berges:2000ew,Litim:2002cf,Bervillier:2007rc}.

For our fixed-point analysis, we study the first derivative of Eq.~\eqref{eq:fixPoints3},
\begin{equation}
  	\partial_t w_k'=\frac{\varphi w_k''-3w'_k}2+
		\frac{w_k''^{\,2}w_k'''}{4\pi^2(1+w_k''^{\,2})^2}-
		\frac{w_k'''}{8\pi^2(1+w_k''^{\,2})},
		\label{eq:fixPoints5}
\end{equation}
where the prime denotes the derivative with respect to the dimensionless field $\varphi$.
\subsubsection{Polynomial approximation}
First we solve Eq. \eqref{eq:fixPoints5} 
in the polynomial approximation with a $\Z_2$ symmetric
$w_\Lambda'$ at the cutoff scale. The RG flow is such that a
$\Z_2$ symmetric $w'_\Lambda$ will remain $\Z_2$ symmetric during the flow. 
Thus a polynomial approximation for $w_\Lambda'$ is of the form
\begin{equation}
  w_k'(\varphi)=\lambda(t)\left(\varphi^2-a^2(t)\right)+\sum_{i=2}^n b_{2i}(t)\,{\varphi}^{2i}\,,
\label{eq:polexpans}
\end{equation}
where $\lambda$, $a^2$ and $b_{2i}$ denote the scale-dependent
couplings. Recall that for an even function $w'(\varphi)$ supersymmetry 
may be broken. We find the following infinite tower of differential equations:
\begin{align}
\nonumber
\partial_ta^2(t)&=a^2(t) \left(-\frac{3 \lambda(t){}^2}{\pi
   ^2}+\frac{3 b_4(t)}{2\pi ^2
   \lambda(t)}-1\right)+\frac{1}{4\pi ^2}\,,\label{eq:a2flow}\\
\partial_t\lambda(t)&=-\frac{3 b_4(t)-6 \lambda (t)^3+\pi ^2 \lambda (t)}{2 \pi
   ^2}\,,\\
 \nonumber
\partial_t b_4(t)&=\frac{120 b_4(t) \lambda (t)^2+2 \pi ^2 b_4(t)-15
   b_6(t)-80 \lambda (t)^5}{4 \pi ^2}\\
   \vdots\nonumber
 \end{align}
Note that due to supersymmetry the lowest order coupling $a^2$ does not enter 
the flow equations of the couplings $\lambda,b_4, b_6,\dots$. 

In our fixed-point analysis we find a Gau{\ss}ian fixed point with 
all coupling constants equal to zero and, due to the $\mathbb Z_2$ symmetry, a
pair of fixed-points whose couplings converge rapidly for larger truncations as shown 
in Tabular~\ref{tab:FPCoefficients}.
From the stability matrix,
\begin{equation}
B_i{ }^j=\frac{\partial{(\partial_t b_i)}}{{\partial b_j}}\,,
\end{equation}
we read off that the non-Gau\ss ian fixed points are IR stable. Here,
we have set $b_0=a^2$ and $b_2=\lambda$. These IR stable fixed points are to be considered as 
supersymmetric equivalent of the  Wilson-Fisher fixed point.

\begin{table}
\begin{center}
\begin{footnotesize} 
\begin{ruledtabular}
\begin{tabular}{ccccccc}
$2n$&$\pm \lambda$&$\pm b_4$&$\pm b_6$&$\pm b_8$&$\pm b_{10}$&$\pm
b_{12}$\\\hline 4 &1.546& 2.305\\
6&1.590&2.808&6.286\\
8& 1.595&2.873& 7.150& 13.41\\
10 & 1.595& 2.873& 7.155&13.48& 1.212\\
12& 1.595& 2.870& 7.118& 12.90&-8.895& -183.3
\end{tabular}
\end{ruledtabular}
\caption{Wilson-Fisher fixed point as obtained from the polynomial approximation
to $w'(\varphi)$. Note that the discrepancy between the fixed-point
values of $b_{10}$ for the truncation with $2n=10$ and $2n=12$ is a truncation effect. 
Due to our polynomial expansion of $w_k'(\varphi)$, see Eq.~\eqref{eq:polexpans}, 
we obtain a hierachy of flow equations for the couplings $b_{2i}$ in which the RG flow of 
the coupling $b_{2n}$ (for given truncation order $n$) suffers the most by the finite 
truncation order $n$. Therefore we expect that the fixed-point values for the higher-order 
couplings converge when $n$ is increased, as it is indeed the case for $\lambda$, $b_2$, $b_4$, $b_6$.
\label{tab:FPCoefficients}}
\end{footnotesize}
\end{center}
\end{table}

Let us now discuss the critical exponents which are the negative 
eigenvalues of the stability matrix at the fixed point. The coupling $a^2$, which 
does not feed back into the equations for the higher-orders couplings,
defines an IR unstable direction with a critical exponent $1/{\nu_{\rm { }_W}}{ =\frac 32}$. 
The critical exponents of the IR stable directions of the Wilson-Fisher fixed point
are given in Tabular~\ref{tab:criticalExponents1}. Actually we observe a better 
convergence of the lowest critical exponents than in two dimensions~\cite{Synatschke:2009nm}.

\begin{table}
\begin{center}
\begin{footnotesize} 
\begin{ruledtabular}
\begin{tabular}{cccccccccc}
 $2n$ &\multicolumn{9}{c}{critical exponents} \\\hline
$6$&$-0.799$&$ -5.92$&$ -20.9$\\
$8$&$-0.767$&$ -4.83$&$ -14.4$&$- 38.2$\\
$10$&$-0.757$&$ -4.35$&$ -11.5$&$- 26.9$&$ -60.8$\\
$12$&$-0.756$&$ -4.16$&$ -9.94$&$ -21.4$&$ -43.8$&$ -89.0$\\
$14$&$-0.756$&$ -4.10$&$ -9.13$&$ -18.3$&$- 35.1$&$ -65.4$&$-
123$\\
$16$&$-0.756$&$- 4.08$&$ -8.72$&$ -16.4$&$ -29.9$&$ -52.9$&$ -91.9$&$ 
-163$\\
$18$&$-0.756$&$- 4.08$&$ -8.54$&$ -15.2$&$ -26.4$&$- 45.0$&$
-75.0$&$  -124$&$ -209	$
\end{tabular}
\end{ruledtabular}
\caption{Critical exponents for the Wilson-Fisher fixed point for different
truncations.
\label{tab:criticalExponents1}}
\end{footnotesize}
\end{center}
\end{table}

\begin{figure}[t]
\includegraphics[width=\columnwidth]{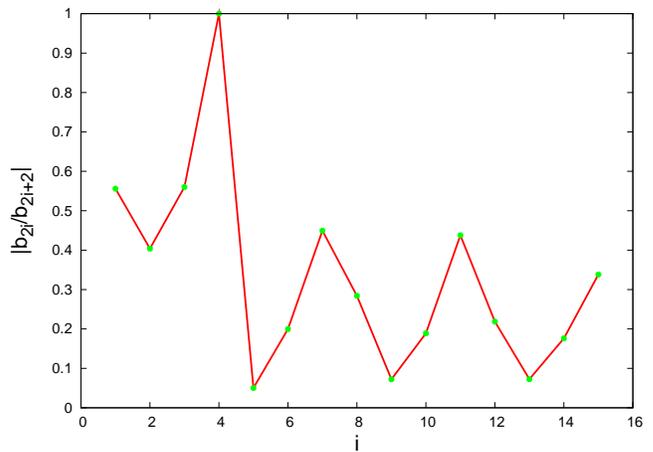}
\caption{Ratio $b_{2i}/b_{2i+2}$ of the Taylor expansion coefficients of the potential as a function of $i$.
\label{fig:finiteRadofConv}}
\end{figure} 
From Tabular~\ref{tab:FPCoefficients} we estimate that the 
radius of convergence of the Taylor series \eqref{eq:polexpans}, given by 
\begin{equation}
  r=\lim_{i\to\infty}\frac{b_{2i}}{b_{2i+2}},
\end{equation}
is finite. This can be seen more clearly
from Fig.~\ref{fig:finiteRadofConv}, where we plotted the ratios $b_{2i}/b_{2i+2}$ as functions of $i$.
Note, that for the corresponding series 
with dimensionful field and couplings
$\bar b_{2i}=b_{2i}\cdot k^{3/2-i}$ the radius of convergence shrinks with decreasing scale $k$,
\begin{equation}
  \bar r=\lim_{i\to\infty}\frac{\bar b_{2i}}{\bar
  b_{2i+2}}=k\lim_{i\to\infty}\frac{ b_{2i}}{
  b_{2i+2}}=k \cdot r\,.
\end{equation}
Thus, the radius of convergence of the power series expansion of $W'_k(\phi)$ 
tends to zero for $k\to 0$. 

\subsubsection{Partial differential equation}
Let us now turn to the solution of the partial differential equation
\eqref{eq:fixPoints3}. We have seen in Eq.~\eqref{eq:a2flow} that
the  coupling associated with the IR unstable direction does not feed back
into the fixed-point equation. It is therefore sufficient to consider the second
derivative of Eq.~\eqref{eq:fixPoints3}. To simplify the
notation we introduce $u(\varphi)=w''_k(\varphi)$. The fixed-point 
equation for $u$  reads
\begin{equation}
u''
=2u\,\frac{u^2-3}{u^4-1}\,u^{\prime\,2}
+4\pi^2\frac{(u^2+1)^2}{u^2-1}(2u-\varphi u').
\end{equation}
It is straightforward to see that the above equation has an 
asymptotic solution $u_{\rm as}(\varphi)\sim\varphi^2$. Employing
a standard Runge-Kutta solver for ordinary differential equations
we find one \emph{regular odd solution} for the starting condition 
$
u(0)=0$ and $
u'(0)\equiv2\lambda\approx\pm 2\cdot
1.59508=\pm 3.19016\,.
$ 
For field amplitudes $\varphi<0.245$ the regular solution is
bounded by $u<1$ and this inner part of the solution corresponds to the IR stable fixed-point solution found in the polynomial approximation discussed above. 

Since $u=w''$ we find for large fields 
\begin{equation*}
w''(\varphi\to\pm \infty)\simeq \pm\,\varphi^2
\Longrightarrow W''_\ast(\phi\to\pm\infty)\simeq \pm\,\phi^2.
\end{equation*}
In other words the outer part of the regular fixed-point solution connects 
smoothly to the inner part. Thus we have found a solution corresponding to a bosonic potential
$V\simeq (k^{3/2}w'_\ast(\varphi))^2$ which behaves as $\sim\phi^6$ for large
$\phi$. This is the supersymmetric analogue of the Wilson-Fisher fixed point 
of three-dimensional $O(N)$ theories, see
e.\,g.~\cite{Tetradis:1993ts,Berges:2000ew}.

\subsection{Next-to-leading order}\label{sec:NTLO}

\begin{figure*}
\includegraphics[width=\columnwidth]{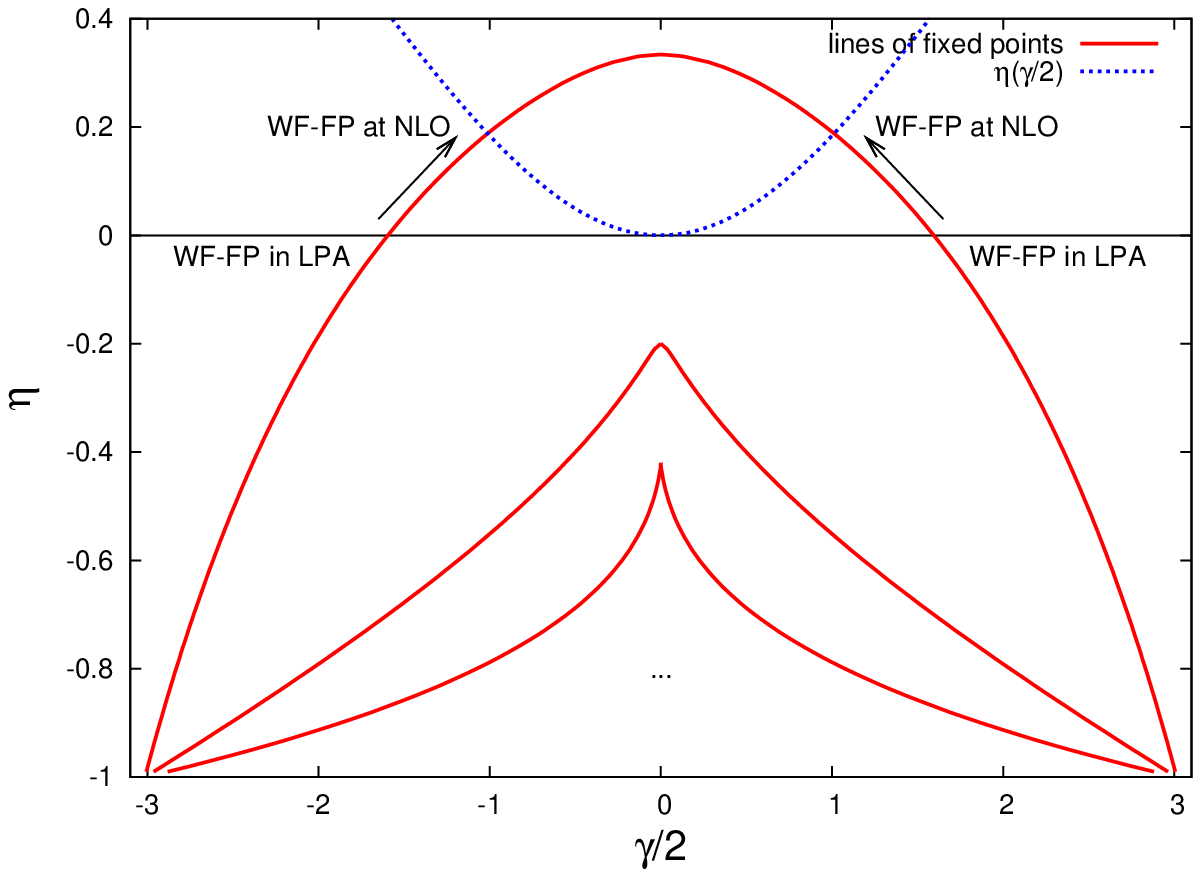}
\includegraphics[width=\columnwidth]{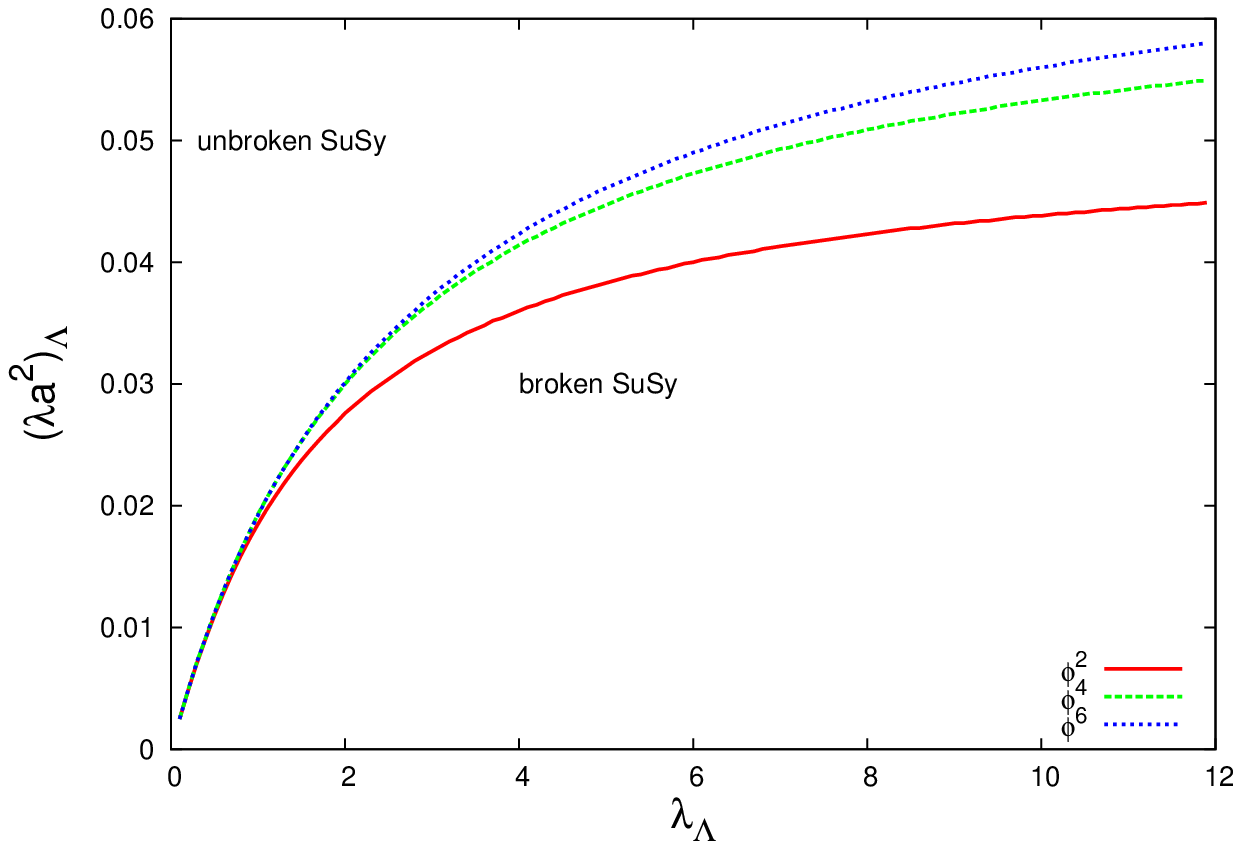}
\caption{\emph{Left panel:} Lines of fixed points in the
	$\eta$-$\gamma$ plane (solid curves) and the anomalous dimension as a 
	function of $\gamma=2\lambda$ as obtained from Eq. \eqref{eq:NextToLead4} 
	(dotted curve). 
	\emph{Right panel:} Phase diagram in the plane spanned by the
dimensionless couplings specified at the cutoff scale  $\Lambda$ as obtained from 
truncations with $n=1$ ($\phi^2$), $n=2$ ($\phi^4$), and $n=3$ ($\phi^6$) in 
Eq.~\eqref{eq:polexpans}.
\label{fig:Phasediagram}}
\end{figure*}
For the next-to-leading order approximation we employ the following ansatz
\begin{equation}
  \begin{split}
 \Gamma_k
 =\int d^3x\left(\frac12 Z_k^2\left(\partial_\mu\phi\partial^\mu\phi-
  	i\psib\slashed\partial\psi
  	 F^2\right)\right.\\\left.+FW_k'(\phi)-\frac12W_k''(\phi)\psib\psi\right)\,,
  	\end{split}
\end{equation} 
with $Z_k^2$ being a scale-dependent wave-function renormalization. 
We neglect a possible momentum and $\phi$ dependence of $Z_k$. This approximation
corresponds to an inclusion of the next-to-leading order correction\footnote{LPA corresponds
to the leading-order approximation.}
in a systematic expansion of the effective action in powers of fields and derivatives. As we discuss below, 
we find that the anomalous dimension $\eta$ remains small compared to one within this
approximation, see also Tabular~\ref{tab:anomalousDimension}. 
Thus, we expect that higher-order corrections such as $\sim (\phi\partial_{\mu}\phi)^2$ does
not significantly affect our results for the zero- and finite-temperature phase 
diagram\footnote{Note that the same reasoning has been found to hold in studies
of the critical dynamics in $O(N)$ models and low-energy QCD models, see 
e.~g. Refs.~\cite{Tetradis:1993ts,Berges:2000ew,Bohr:2000gp,Benitez:2009xg,Braun:2009si}}. 

Projecting on the part linear in the auxiliary field  and integrating with respect to 
$\phi$ yields the superpotential. On the other hand projecting on the terms 
quadratic in the auxiliary fields yields the flow equation for the wave function renormalization.
Employing the regulator functions \eqref{eq:lpa5}  we find the following 
coupled set of differential equations:
\begin{align}
\partial_kW_k(\phi)=&-\frac{k^2 W''_k(\phi)}{24 \pi ^2 }\,
\frac{k \partial_kZ_k^2+3 Z_k^2}{k^2 Z_k^4+W''_k(\phi)^2}\,,\\
\nonumber 
\partial_kZ_k^2=&-\frac{k^2}{4 \pi ^2 } (k \partial_kZ_k^2+2Z_k^2)\times
\\&\left.\times\frac{ Z_k^2 W_k^{(3)}(\phi )^2 \left(k^2 Z_k^4-W_k''(\phi
)^2\right)}{\left(k^2 Z_k^4+W_k''(\phi )^2\right)^3}\right|_{\phi=0}\,.
\end{align}
Introducing the anomalous dimension $\eta=-\partial_t\ln Z_k^2$
and the dimensionless quantities 
\[
\chi=Z_kk^{-\frac12}\phi,\quad\w(\chi)=k^{-2}W_k(\phi),\quad
\]
the dimensionless flow equations read
\begin{align}
	\partial_t\w+2\w&=\frac12(1+\eta)\chi\,\w
		-\frac{(3-\eta)\w''}{24\pi^2(1+\w''^2)}\,,
		\\
	\eta&=\left.\frac{(2\!-\!\eta)(1\!-\!\w''^2)\w'''^2}{4\pi^2(1\!+\!\w''^2)^3}
		\right|_{\chi=0}\!\!\!\!\!\!\!\!\!\!.\label{eq:NextToLead4}
\end{align}
In order to study the fixed-point structure it is convenient to consider the
anomalous dimension $\eta$ as a free parameter\footnote{Note that $\eta=0$ is a consistent
solution of the fixed-point equations. In this respect the model in three 
dimensions is substantially different from the model in two dimensions. 
There, $\eta=0$ is not a consistent solution of the fixed-point equations.}, 
see e.~g. Ref.~\cite{Neves:1998tg}. In complete analogy to the two-dimensional Wess-Zumino 
model we find lines of fixed points corresponding to potentials with no nodes (outermost line),
one node and so on, see Fig.~\ref{fig:Phasediagram} (left panel). In fact, we encounter 
exactly the same picture as in two dimensions, apart from a shift of the graph
to lower $\eta$ values. Concerning the number of fixed points the situation is
completely different as in two dimensions: Because of the shift of $\eta$ we 
find only \emph{one pair} of fixed points for $\eta=0$, and not an infinite
number of pairs. Such a dependence on the dimensionality has also been observed 
for $O(N)$ models~\cite{Neves:1998tg}.

Note that if we had actually used the same regulator as in \cite{Synatschke:2009nm}, namely
\begin{equation*}
r_1=0\quad\hbox{and}\quad r_2=\left(\frac{k^ 2}{p^2}-1\right)\theta(k^2-p^2),
\end{equation*} 
then the flow equation for the superpotential in two dimensions would turn into
the flow equation in three dimensions under the transformation $\eta\to\eta-1$. 
This correspondence explains the similarities of Fig.~\ref{fig:Phasediagram} (left panel) in
two and three dimensions.

As in two dimensions, we can deduce a \emph{superscaling relation} from the RG flow 
equation of $a^2$ which relates the critical exponent $1/{\nu_{\rm { }_W}}$ and the anomalous 
dimension~\cite{Gies:2009az}:
\begin{equation}
  \frac{1}{\nu_{\rm {}_W}}=\frac{3-\eta}2\,.
\end{equation} 
The truncation dependence of the fixed-point value of the anomalous dimension is shown 
in Tabular~\ref{tab:anomalousDimension}.

\begin{table}
\begin{center}
\begin{ruledtabular}
\begin{tabular}{cccccccc}
 $2n$&4&6&8&10&12&14\\\hline
 $\eta_*$&	0.187711&0.188258&0.18802&0.187996&0.188001&0.188003
\end{tabular}
\end{ruledtabular}  	
\caption{Dependence of the fixed-point value of the anomalous dimension $\eta$ on the 
truncation order $n$ of the expansion of $w'$ in powers of the field $\varphi$, see also
Eq.~\eqref{eq:polexpans}.
\label{tab:anomalousDimension}}
\end{center}	
\end{table}

\subsection{Zero-temperature phase diagram and scaling of the mass}
%
\begin{figure}
\includegraphics[width=\columnwidth]{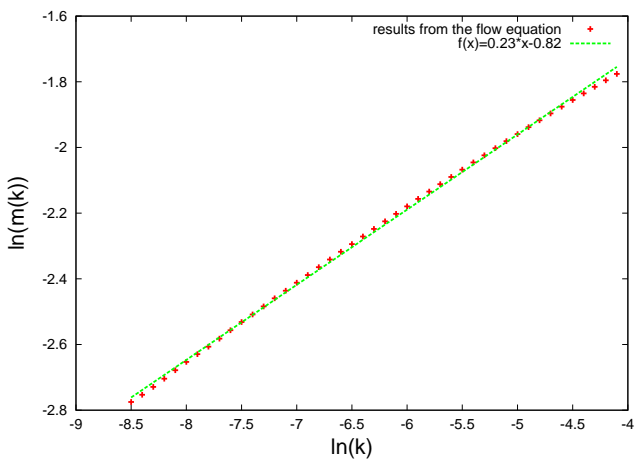}
\caption{Logarithm of the boson mass as a function of the RG-scale $k$. A
linear fit to the data points yields $m(k)\sim k^{0.23}$ for $k\ll \Lambda$.
\label{fig:MassVersusK}}
\end{figure}

The breaking of supersymmetry is driven by the unstable direction $a^2$,
similarly as in two dimensions~\cite{Synatschke:2009nm}.
In the plane spanned by the values of the dimensionless couplings 
$\lam$ and $\lam a^2$ given at the cutoff-scale $\Lambda$ we find a transition 
line for supersymmetry breaking, see Fig.~\ref{fig:Phasediagram} (right panel).

From the effective potential  $V=\lim_{k\to0}W'^2_{k}/2$, we read off the
mass of the boson,
\begin{equation}
  	m^2_{\rm bos}=W'_{k\to 0}(\phi_0)\, W'''_{k\to 0}(\phi_0)+
W''^{\,2}_{k\to 0}(\phi_0)\,,
\end{equation}
where the field $\phi_0$ minimizes $V$.
In the case of broken supersymmetry (and unbroken $\Z_2)$ we have $\phi_0=0$ 
and the mass is given by
\begin{equation}
  	m^2_{{\rm broken}}
  	=W'_{k\to 0}(0)W'''_{k\to 0}(0),\quad W''_{k\to 0}(0)=0,
\end{equation}
whereas for unbroken supersymmetry it is given by
\begin{equation}
  	m_{\rm unbroken}^2=W''^{\,2}_{k\to 0}(\phi_0),\quad W'_{k\to 0}(\phi_0)=0\,.
\end{equation} 
In the broken phase a polynomial expansion of the superpotential is 
justified\footnote{Note that for unbroken supersymmetry a polynomial expansion is
bound to fail and one needs to solve the partial differential
equation in order to determine the boson mass.} and we find
\begin{equation}
m_{{\rm broken}}^2 (k)=
	W_k'(0) W_k'''(0) = 2 k^2\lambda^2 a^2\,.
\end{equation}
From the  scaling of the couplings for  $k\to0$,
\begin{equation}
\lambda_\ast\sim \text{const.}\quad\text{and} \quad
	a^2\sim k^{-\frac32}\,,
\end{equation}
we read off that the mass scales as
\begin{equation}
m(k)\sim k^{\frac14}\quad\hbox{for}\quad k\ll\Lambda\,.\label{eq:MassScaling}
\end{equation}
This scaling behavior is demonstrated in Fig.~\ref{fig:MassVersusK} for a truncation
with $n=4$, i. e. an expansion of $w'$ up to order $\varphi^8\sim \phi^8$, 
see also Eq.~\eqref{eq:polexpans}.
Due to our analysis of the convergence of the fixed-point
values of the couplings $\lambda$ and $b_{2i}$ as well as the anomalous dimension $\eta_*$, 
see Tabular~\ref{tab:FPCoefficients}, \ref{tab:criticalExponents1} and~\ref{tab:anomalousDimension}, 
we expect that the truncation order $n=4$ is already sufficient to capture
qualitatively and quantitatively most of the features of the zero- and finite-temperature
phase diagram, see also Sect.~\ref{sec:highT}. 
From a linear fit to the double-logarithmic plot of $m(k)$ we find $m(k)\sim k^{0.23}$ 
which is indeed reasonably close to the expected scaling given in Eq.~\eqref{eq:MassScaling}. 
The reason for this behavior, which is very different from the one found 
in $O(N)$ models, is that the unstable direction does not feed back into the fixed-point 
equation\footnote{Note that this is only true for finite $N$. In the large $N$ limit,
the running of the higher-order couplings is also independent of the running of the
vacuum expectation value of the field~\cite{Tetradis:1993ts} which corresponds to $a^2$ in our
study of the ${\mathcal N}=1$ Wess-Zumino model.}. 
Independent of the value of the coupling at the UV (ultraviolet) cutoff scale $\Lambda$, the second derivative 
of the superpotential flows always into its IR fixed point corresponding to
a conformally invariant theory. In $O(N)$ models, on the other hand, 
the unstable direction in the broken regime feeds back
into the fixed-point equations for the couplings. Approaching the IR fixed point 
of $O(N)$ models requires therefore fine tuning of the UV parameters~\cite{Tetradis:1993ts}.

\section{Flow equations at finite temperature}\label{sec:FEAFT}
In this section we study finite temperature effects in the three-dimensional Wess-Zumino 
model. To this end we restrict ourselves to the LPA which we expect to provide already 
a quantitative insight into the finite-temperature phase structure as shown in Ref.~\cite{Braun:2009si}.

The finite-temperature flow equation in LPA can be obtained straightforwardly from the zero-temperature 
equations \eqref{eq:lpa7} by replacing the momentum integration in time-like direction by
a summation over Matsubara frequencies: 
\begin{equation}
p_0\longrightarrow \left\{ \begin{array}{c} \nu_n \\ \omega_n \end{array} \right\}\,,
\quad \int\frac{dp_0}{2\pi}\dots \longrightarrow T\sum_{n=-\infty}^\infty\dots
\end{equation} 
with frequencies $\omega_n=2\pi nT$ for bosonic fields and $\nu_n=(2n+1)\pi T$ for
fermionic fields. We refer to Appendix~\ref{sec:FiniteTemp} for a detailed 
derivation of the finite-temperature flow equations. Here we simply note that
we can perform the Matsubara sums explicitly for the regulator functions
\eqref{eq:lpa4}. The flow equations read
\begin{align}
\partial_k{W'_k}^{\rm bos}
	=&-\frac{k^2}{8\pi^2 }W_k'''\frac{k^2-W''^2_k}{(k^2+W''^2_k)^2}\times\\&
	\nonumber
	\left( \frac{\pi T}{k}
          -(2s_{\rm B}+1)^2\frac{\pi T}{k}
          +2(2s_{\rm B}+1)
	\right)\frac{\pi T}{k}\,,\\
\partial_k{W'_k}^{\rm ferm}
	=&-\frac{k^2}{8\pi^2 }\frac{(k^2-W''^2_k)W_k'''}{(k^2+W''^2_k)^2}
	\biggl(\!1-\Bigl(1-\frac{2s_{\rm F}\pi T}{k}\Bigr)^2 \biggr),
\end{align}
where the \emph{temperature-dependent} floor functions $s_{\rm B}$ and $s_{\rm F}$ are given by
\begin{equation}
s_{\rm B}=\left\lfloor\frac{k}{2\pi T}\right\rfloor\quad\text{and}\quad
s_{\rm F}=\left\lfloor\frac{k}{2\pi T}+\frac12\right\rfloor\,.\label{floor:fktn}
\end{equation}
The differences in the flow equations for the 'superpotential' describing the 
self-interaction of bosons and the 'superpotential' describing the Yukawa-type interaction 
between fermions and bosons originates in the supersymmetry breaking induced by the 
different thermal boundary conditions for the bosonic and fermionic fields.

\subsection{Pressure}\label{sec:pressure}

\begin{figure}
\includegraphics[width=\columnwidth]{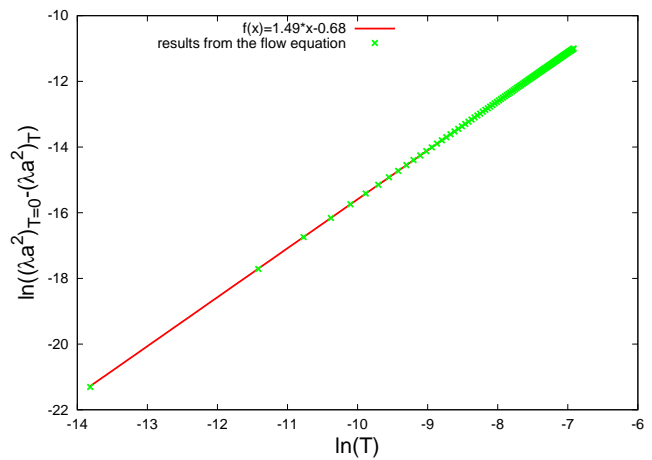}
\caption{Double logarithmic plot of the temperature dependent minimum of the
superpotential versus the temperature. \label{fig:pressure}}
\end{figure}
In the previous section we have shown that the boson mass tends to 
zero for $k\to 0$ in the phase with broken supersymmetry (restored $\Z_2$ symmetry). 
Thus we expect that the thermodynamic properties of the three-dimensional model in the 
phase with restored $\Z_2$ symmetry are similar to that of a gas of massless
bosons. The pressure of such a gas in $2+1$ dimensions is given by
\begin{equation}
  \Delta p=\frac{\zeta(3)}{2\pi}\,T^3.
\end{equation}
The (normalized) pressure $\Delta p$ for a given temperature is determined
by the negated difference of the interaction potential for a given 
temperature evaluated at the minimum and its corresponding zero-temperature value. In our model 
this corresponds to the height of the barrier in the 
double-well potential $(W')^2$ for a theory with broken $\mathbb Z_2$
symmetry or the minimum of $(W')^2$ with unbroken $\mathbb Z_2$
symmetry. For small temperatures we therefore expect 
\begin{align}
\nonumber
 \Delta(\lambda a^2)\equiv&(\lambda a^2)_{T=0}-(\lambda
  a^2)_{T\neq0}\\=&\sqrt{\frac{\zeta(3)}{2\pi}}\cdot T^{\frac32}\simeq
  \exp(-0.83) T^{\frac32}\,
  \label{eq:Pressure3}
\end{align}
for the temperature dependence of the minimum of $W'(\phi)$.
In Fig.~\ref{fig:pressure} we show the double-logarithmic plot 
of the minimum of the superpotential versus temperature as obtained from a truncation
of the potential with $n=4$, i. e. an expansion of $w'$ up to order $\varphi^8 \sim \phi^8$, 
see also Eq.~\eqref{eq:polexpans}. 
The linear fit to the double-logarithmic plot yields a power law $\Delta(\lambda a^2)=\exp(-0.68)\cdot
T^{1.49}$  which is compatible with the values for a gas of massless bosons
in $2+1$ dimensions, as given in Eq.~\eqref{eq:Pressure3}. 
We expect that deviations from the ideal gas limit are present for 
two reasons: (i) The boson mass tends to zero for $k\to 0$ only but remains finite for any $k>0$, 
see Eq.~\eqref{eq:MassScaling} and (ii) boson self-interactions lead to deviations 
from the ideal bose-gas limit.

\subsection{RG flows at finite temperature and dimensional reduction}\label{sec:highT}
For high temperatures $T\gg k$ both floor functions in \eqref{floor:fktn} vanish
and the flow equations simplify considerably: 
\begin{align}
\partial_k{W'_k}^{\rm bos}
	&=-\frac{k^2}{8\pi^2 }W_k'''\frac{k^2-W''^2_k}{(k^2+W''^2_k)^2}
	\frac{2\pi T}{k}\,, \label{eq:FlowTemp1}\\
\partial_k{W'_k}^{\rm ferm}
	&= 0\,.\label{eq:FlowTemp3}
\end{align}
Following Ref.~\cite{Tetradis:1992xd}  we rescale the quantities in the bosonic
flow equation according to
\begin{eqnarray}
  	\phi&=&\sqrt{T}\tilde\phi,\quad W_k(\phi)=T\tilde W_k(\tilde\phi)\,.
	\label{eq:TempResc}
\end{eqnarray}
For Eq. \eqref{eq:FlowTemp1} this leads exactly to the zero-temperature flow
equation of the two-dimensional model. At this point, however, we should stress that the
theory which we obtain in the limit $T/k \gg 1$ is \emph{not} the supersymmetric
$\N=1$ Wess-Zumino model in two dimensions since the fermions have dropped out
of the theory due to the absence of thermal zero-modes. Therefore supersymmetry is
necessarily broken at finite temperature. The fact that we still obtain the 
same functional form for the bosonic flow equation can be understood in terms of the role
of the auxiliary field: In order to obtain the bosonic flow equation, we have to project on the terms 
coupling to the auxiliary field. Since there is no coupling between the auxiliary field and 
the fermionic part of the theory, the fermions do not contribute to the bosonic flow equation.

Because of the rescalings of the fields and the potential according to 
Eq.~\eqref{eq:TempResc} we expect that the couplings exhibit the following behavior for $T/k\gg 1$: 
\begin{align}
\nonumber
{^{3D}}(a^2)_{T}&={^{2D}(a^2)_{T=0}}\left(\frac Tk\right)^{1/2},\\
{^{3D}}\lambda_{T}&={^{2D}}\lambda_{T=0}\left(\frac Tk\right)^{-1/2},\\
{^{3D}}(b_{2i})_{T}&={^{2D}}(b_{2i})_{T=0}\left(\frac
Tk\right)^{1/2-i},\nonumber
\label{eq:2d3drunning}
\end{align}
where ${^{2D}(a^2)_{T=0}},\, {^{2D}}\lambda_{T=0}$ and ${^{2D}}(b_{2i})_{T=0}$ denote the fixed-point values
of the couplings of the two-dimensional theory. Indeed we observe such a running of the couplings for $T/k\gg 1$ in 
our numerical studies.

From Eq.~\eqref{eq:2d3drunning} we deduce the radius of convergence 
of our expansion of the potential in powers of the fields. Recalling the
relation between the dimensionful and dimensionless couplings,
$\bar b_{2i}=b_{2i}\cdot k^{\,3/2-i}$, we find
\begin{align}
 ^{3D}r=&\lim_{i\to\infty}\frac{ ^{3D}{(\bar b_{2i})_{T}}}{^{3D}{(\bar
 b_{2i+2})_{T}}}\nonumber\\
 =&\lim_{i\to\infty}\frac{^{2D}{(b_{2i})_{T=0}}\cdot(T/k)^{1/2-i}\cdot
 k^{\,3/2-i}} {^{2D}{(b_{2i+2})_{T= 0}}\cdot(T/k)^{-1/2-i}\cdot
 k^{\,1/2-i}}\\
 =&{^{2D}}r\cdot T\,. \nonumber
\end{align} 
Thus we find a finite radius of convergence for the polynomial expansion of 
the superpotential at finite temperature which is a consequence of the finite 
radius of convergence found for the underlying two-dimensional 
model at zero temperature~\cite{Synatschke:2009nm}. However, the radius of 
convergence $^{3D}r$ tends to zero for $T\to 0$, in full agreement with our results 
in Sect.~\ref{subsec:LPA}.

\subsection{Phase diagram at finite temperature}\label{sec:PTPD}
%
\begin{figure*}
\includegraphics[scale=0.7]{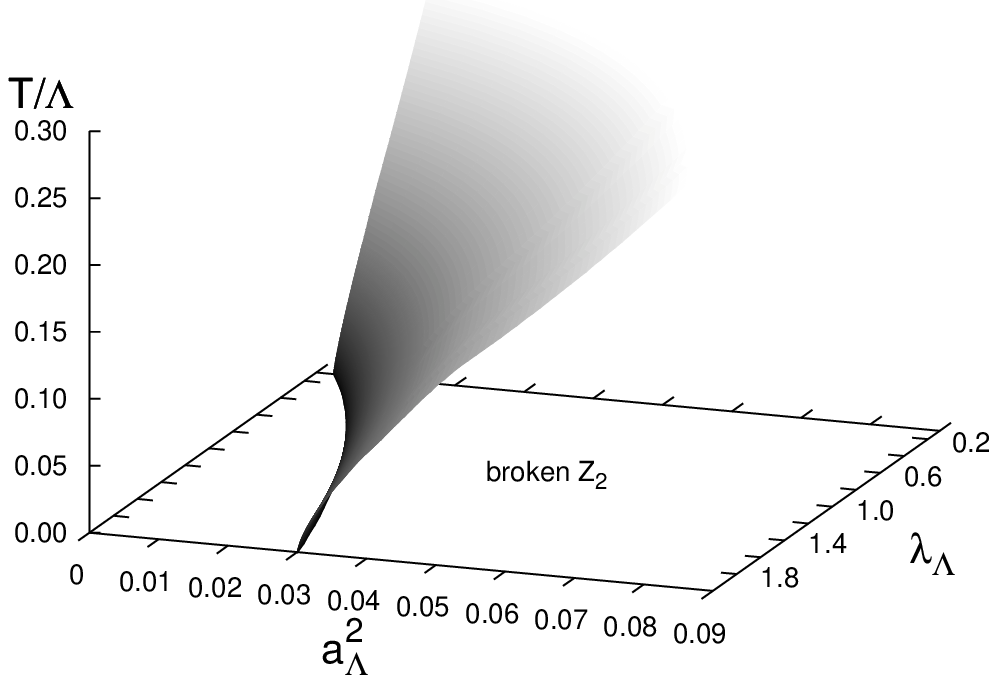}
\includegraphics[scale=0.55]{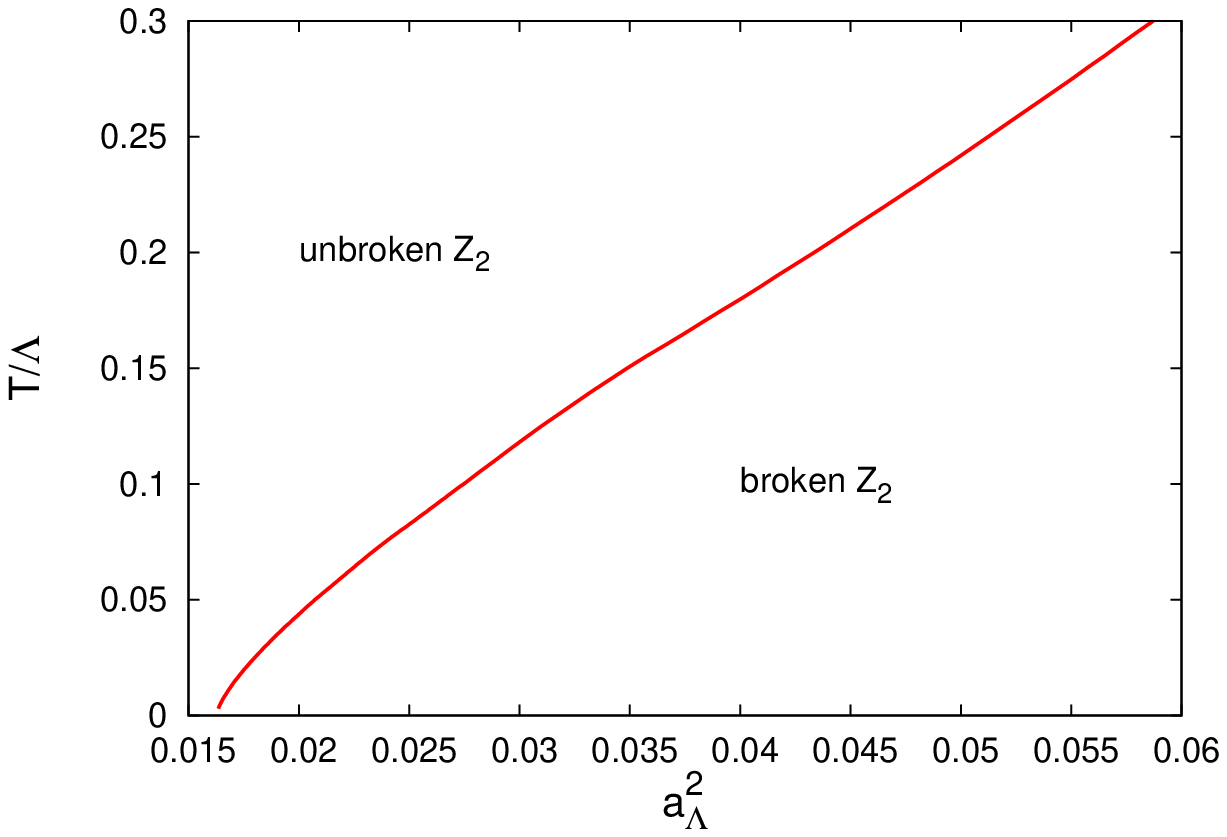}
\caption{Finite-temperature phase diagram of the ${\mathcal N}=1$ Wess-Zumino
model. \emph{Left panel:} $\Z_2$ phase boundary in the space spanned by
temperature $T/\Lambda$ and the value of the couplings $(\lam_{\Lambda},a^2_{\Lambda})$ 
at vanishing temperature. \emph{Right panel:} Slice of the $\Z_2$ phase-boundary
for fixed $\lam_{\Lambda}=0.8$.\label{fig:FiniteTempPhaseDiagram}}
\end{figure*} 
In this subsection we discuss the phase diagram of the ${\mathcal N}=1$ Wess-Zumino
model at finite temperature. 

Whether supersymmetry is broken or not at vanishing 
temperature depends on our choice for the couplings at the cutoff scale as we
have discussed above. At finite temperature we have necessarily soft supersymmetry
breaking due to the different boundary conditions for bosons and fermions in
Euclidean time direction. Besides supersymmetry, however, our theory is invariant under
$\phi \to -\phi$. At vanishing temperature the breakdown of this ${\mathbb Z}_2$
symmetry is intimately linked to the question whether supersymmetry is broken or not.
Recall that at $T=0$ broken $\Z_2$-symmetry of the ground state implies a supersymmetric
ground state and that restored $\Z_2$ symmetry of the ground state implies broken supersymmetry.
Even though supersymmetry is necessarily broken at finite temperature, we shall see
in the following that $\Z_2$ symmetry of the ground state can be either broken or restored 
depending on the actual value of the temperature. Due to the relation between supersymmetry 
and ${\mathbb Z}_2$ symmetry at vanishing temperature, we consider the strength of ${\mathbb Z}_2$
breaking as a measure for supersymmetry breaking at finite temperature. Thus we distinguish between
the case of soft supersymmetry breaking due to finite temperature but broken $\Z_2$ symmetry 
of the ground state and the case with restored $\Z_2$ symmetry of the ground state at finite temperature.

For our numerical study of the finite-temperature phase diagram we employ a $\phi^8$-truncation 
of the potentials\footnote{The quality of such an approximation has been studied extensively
for zero and finite temperatures~\cite{Tetradis:1993ts,Papp:1999he,Litim:2002cf,Bervillier:2007rc}. 
It indeed proves to be quantitatively useful in case one 
is not interested in a high-accuracy determination of critical exponents~\cite{Litim:2002cf,Bervillier:2007rc}.}, 
i.\,e. we use $n=4$ in Eq.~\eqref{eq:polexpans}.

In the left panel of Fig.~\ref{fig:FiniteTempPhaseDiagram}, we show the ${\mathbb Z}_2$ 
phase-boundary in the space spanned by temperature $T/\Lambda$ and the value of the couplings 
$(\lam_{\Lambda},a^2_{\Lambda})$ at the UV cutoff scale at vanishing 
temperature\footnote{As initial conditions for the finite-temperature RG flows we use the 
zero-temperature values of the couplings $(\lam_{\Lambda},a^2_{\Lambda})$ at the cutoff scale.
Therefore we have to restrict our study to temperatures significantly smaller than the UV cutoff 
in order to ensure that it is justified to use these zero-temperature values of the couplings
at the UV starting point of the flow.}. 
The lower end of the phase boundary on the $T=0$ plane in Fig.~\ref{fig:FiniteTempPhaseDiagram} 
corresponds to the phase-transition line which 
separates the phase with broken  supersymmetry from the one associated with a supersymmetric ground state, 
see also Fig.~\ref{fig:Phasediagram} (right panel).
Choosing couplings $(\lam_{\Lambda},a^2_{\Lambda})$ at $T=0$ associated with a supersymmetric
ground state (and broken $\Z_2$ symmetry) we find always a second-order phase-transition temperature 
at which the system enters the phase with restored $\Z_2$ symmetry. 
As discussed in Sect.~\ref{sec:pressure} the thermodynamic properties of this $\Z_2$-symmetric 
phase are similar to the ones of a gas of massless bosons. We expect that 
this finite-temperature phase transition falls into the Ising universality class with 
critical exponents determined by the underlying 2$d$ $\Z_2$ theory, namely by 
the exponents of Onsager's solution of the corresponding lattice spin model. 
However, a quantitative analysis of the critical behavior would require the
inclusion of the anomalous dimension in our studies. This is deferred to 
future work.

In the right panel of Fig.~\ref{fig:FiniteTempPhaseDiagram}, we show a slice of the $\Z_2$
phase boundary for fixed $\lam_{\Lambda}=0.8$. We find that the
phase-transition temperature increases with increasing $(a^2_{\Lambda})_{T=0}$. Since $(a^2_{k\to 0})_{T=0}$ increases with increasing
$(a^2_{\Lambda})_{T=0}$, the phase-transition temperature increases with increasing $(a^2_{k\to 0})_{T=0}$, 
i.\,e. in terms of the renormalized quantity\footnote{Recall that $\lam$ flows
into its fixed-point value $\lam_*$ for $T=0$ independently of the initial condition $\lam_{\Lambda}$. Nevertheless $\lam_{\Lambda}$ can be used to
classify different theories at finite temperature.}. Our observation of an increasing phase-transition
temperature with increasing $(a^2_{k\to 0})_{T=0}$ is in accordance with our expectations 
from a scalar $O(1)\simeq\Z_2$ model since $(\lambda_* a^2_{k\to 0})_{T=0}$ sets the scale at $T=0$ and 
therefore plays a similar role as a finite expectation value of the fields in $O(N)$ models, 
see e.\,g.~\cite{Bohr:2000gp}. 

\section{Conclusions}

In this work we have employed the functional RG
for a study of the three dimensional $\N=1$ Wess-Zumino model at zero and at
finite temperature. Since the model exists only in Minkowski space we have worked with 
a formulation of the Wetterich equation in Minkowski-space and have Wick-rotated the 
momentum integrals. 

At zero temperature we have found results quite similar to our
findings for the two-dimensional model~\cite{Synatschke:2009nm}. 
An investigation of the fixed point structure 
yields the supersymmetric analogue of the Wilson-Fisher fixed-point for bosonic theories. 
It is maximally IR stable with one relevant direction only. As in the two-dimensional 
model the relevant direction is given by $a^2$. Again we find a scaling relation 
between the critical exponent $1/\nu_{\rm { }_W}$ of the instable direction and the anomalous 
dimension $\eta$. The critical exponent $\nu_{\rm {}_W}$  governs the freeze-out of the minimum 
of the superpotential $W$. We also find that the IR limit of the three-dimensional 
Wess-Zumino model is given by a conformally invariant theory. The critical exponent $\nu_{\rm {}_W}$ governs
the vanishing of the mass with the RG scale. This is different from three-dimensional 
scalar $O(N)$ theories since for the supersymmetric models the relevant coupling does not feed 
back in the flow equations of higher $n$-point functions. As in two dimensions we find that 
supersymmetry breaking is governed by the relevant direction.

The finite-temperature flow equations are obtained by replacing the integration
over the time-like direction by a sum over Matsubara frequencies. As bosons 
and fermions obey different thermal boundary conditions, finite temperatures 
introduce a soft supersymmetry breaking.

In the phase with broken supersymmetry (restored $\Z_2$ symmetry) at zero temperature 
the three-dimensional Wess-Zumino model flows to a massless field theory. In the $\Z_2$ symmetric phase
at finite temperature the model therefore behaves like a three-dimensional gas of massless bosons. 
We have found that this is indeed the case. The small deviations from the ideal-gas law found in our 
numerical studies originate from the self-interaction of the bosons.

At high temperatures, supersymmetry breaking manifests itself in 
the fact that the flow equations for the superpotential, derived from the 
bosonic and the fermionic part of the action, are different. 
We also observe \emph{dimensional reduction} in the way that, after 
a suitable rescaling, the flow equation for the "bosonic" superpotential  
in three  dimensions reduces to the flow equation in two dimensions. 
Due to the absence of a fermionic thermal zero-mode the
fermions do not contribute to the RG flow for small scales $k\ll T$.
We have argued that the radius of convergence for the polynomial expansion of the
superfield interpolates between the values for two and three dimensions as the
temperature is raised. 

Even though supersymmetry is explicitly broken at finite temperature, the $\Z_2$ 
symmetry of the model can be either restored or broken at finite temperature. Whether
$\Z_2$ symmetry is broken or not depends on the temperature (and parameters of the model, 
i.\,e. the initial values of the couplings at the initial RG scale). Since
supersymmetry and $\Z_2$ symmetry are intimately linked we have argued that a study of $\Z_2$ symmetry may
be used to measure the strength of supersymmetry breaking. We have computed the 
phase diagram for the restoration of $\mathbb Z_2$ symmetry at finite temperatures. We find 
two different types of phases which are separated by a second order phase transition: 
one phase with soft supersymmetry breaking due to finite temperature but broken $\Z_2$ symmetry and one with 
restored $\Z_2$ symmetry.

Throughout this paper we have addressed several similarities and differences between scalar $O(N)$ models and the 
${\mathcal N}=1$ Wess-Zumino model at zero and finite temperatures, e.\,g. the
fixed-point structure at zero temperature and the behavior at finite temperature. A detailed exploration of 
both models with respect to their similarities and differences, in particular with respect to the 
large-$N$ limit, is deferred to future work.
  

\acknowledgments{Helpful discussions with H.~Gies and J.~M.~Pawlowski are
gratefully acknowledged. Moreover the authors are grateful to H.~Gies for
useful comments on the manuscript. FS acknowledges support by the  Studienstiftung des deutschen Volkes. This
work has been supported by the DFG-Research Training Group 
''Quantum-and Gravitational Fields'' GRK 1523/1.}

\appendix

\section{Derivation of the flow equations in Minkowski space}
\label{sec:MinkowskiERG}
In this Addendum we derive the Wetterich flow equation in Minkowski space. For the sake of 
simplicity we consider a real scalar field in this appendix. The generalization to other fields, 
such as  fermion or gauge fields, is straightforward.

The generating functional in Minkowski-space is given by
\[
Z[J]=\int\PfadD \varphi\, e^{i(S[\varphi]+(J,\varphi))}\,,
\]
where $J$ denotes the external source and $(J,\varphi)\equiv\int d^d x\,J(x)\varphi(x)$.
The generating functional $W$ for the connected two-point functions,
the so-called Schwinger functional, reads\footnote{The generating
functional $W$ of connected two-point functions should not be confused with the super potential in the main part.}
\[
W[J]=i\ln Z[j]\,.
\]
From this we obtain
\begin{align*}
\frac{\delta}{\delta J}W[J]&=i\frac{\delta}{\delta J}\ln
Z[J]=-\frac{\int\PfadD\phi\, e^{i (S+\int J\phi)}\phi}{\int\PfadD\phi\, e^{i
(S+(J,\varphi))}}\\&=-\phi=-\ew{\varphi}\,.
\end{align*}
The effective action is the Legendre transform of the Schwinger functional,
\begin{align*}
\Gamma[\phi]=-W[J]-(J,\phi)\,,
\end{align*}
where $\phi$ is the mean (classical) field. Using $\frac{\delta}{\delta J}W[J]=-\phi$,
we obtain the equation of motion for the field $\phi$:
\begin{align*}
\functder{\phi}{\Gamma[\phi]}&=
	-\int d^d y\functder{J(y)}{W[J]}\functder{\phi(x)}{J(y)}
	-\int d^d y\functder{\phi(x)}{J(y)}\phi(y)-J(x)\\
	&=-J(x)\,.
\end{align*}
The scale-dependent generating functional is defined as
\begin{align*}
Z_k[J]=&e^{-iW_k[J]}=e^{i\Delta S_k[\functder{J}{}]}Z[J]\\
=&\int\PfadD \varphi\, e^{i(S[\varphi]+\int_x\varphi J+\Delta S_k[\varphi])}
\end{align*}
with 
\[\Delta
S_k[\phi]=\frac12\int\frac{d^4q}{(2\pi)^4}\varphi(-q)R_k(q)\varphi(q)\,.
\]
Here, the momentum-dependent regulator function $R_k$ provides an IR cutoff for 
all modes and has to satisfy three conditions: (i) $R_k(p)|_{p^2/k^2\to 0} >0$ which implements 
the IR regularization, (ii) $R_k(p)|_{k^2/p^2\to 0} =0$ which ensures that the regulator vanishes for $k\to0$, 
(iii) $R_k(p)|_{k\to\Lambda\to\infty}\to\infty$ which ensures that the path integral is dominated
by the stationary point for $k\to\Lambda\to\infty$. 
Different choices for $R_k$ define different RG trajectories manifesting the RG scheme dependence, but the IR
physics  $\Gamma_{k\to 0}\to\Gamma$ should remain invariant provided the truncation captures all
relevant operators for the physical observables under investigation. In turn, a variation of the regulator function may lead to more insight on the truncation dependence of our results.

Next, we define the scale-dependent effective action:
\begin{align*}
\Gamma_k[\phi]=-W_k[J]-\int d^4x\,J\phi-\Delta S_k[\phi]\,.
\end{align*}
In order to properly formulate the flow equations in Minkowski-space we have to take $k^2\sim p_{\mu}p^{\mu}$ 
as flow parameter. Therefore we define $\partial_t=2 k^2\partial_{k^2}$ to be the derivative with respect to RG 'time' $t=\ln k^2$.
Taking the derivative of $\Gamma_k[\phi]$ with respect to $t$ yields
\begin{equation}
\partial_t\Gamma_k[\phi]=-\partial_tW_k[J]-\partial_t\int
d^4xJ\phi-\partial_t\Delta S_k[\phi]\,,\label{eq:GammaFlow1}
\end{equation}
where we have used that the source is independent of $k$. For the derivative of $W_k$ we obtain 
\begin{align*}
&\partial_t W_k[J]=i\partial_t\ln Z[J]\\
&=\frac{i}{Z[J]}\int\PfadD \varphi\frac i2\int\frac{d^d
q}{(2\pi)^d}\varphi\partial_t R_k\varphi e^{i(S[\varphi]+\int_x\varphi J+\Delta S_k[\varphi])}\\
&=-\frac{1}{2Z[J]}\int\frac{d^d q}{(2\pi)^d}\partial_t R_k
	{\int\PfadD \varphi\;\varphi\,\varphi\,
	e^{i(S[\varphi]+\int_x\varphi J+\Delta
	S_k[\varphi])}}\\
&=\frac{1}{2Z[J]}\int \frac{d^d q}{(2\pi)^d}(\partial_t
R_k)\functderder{J}{J}{Z[J]}\,.
\end{align*}
Using the definition of the Schwinger functional, $Z_k[J]=e^{-iW_k}$, we find
\begin{equation}
\partial_t W_k[J]=e^{iW_k}\frac 12\int \frac{d^d q}{(2\pi)^d}(\partial_t
R_k) \functderder{J}{J}{e^{-iW_k}}\,.\label{eq:Wflow}
\end{equation}
Now we rewrite the integrand by making use of 
\begin{align*}
\functderder{J}{J}{e^{-iW_k}}
	&=\functder{J}{}e^{-iW_k}(-i)\functder{J}{W_k}\\
	&=e^{-iW_k}(-i)\functder{J}{W_k}(-i)\functder{J}{W_k}+
	e^{-iW_k}(-i)\functderder{J}{J}{W_k}\,,
\end{align*}
then Eq.~\eqref{eq:Wflow} can be rewritten as follows:
\begin{align*}
\partial_tW_k[J]
	=&\frac 12\int \frac{d^d q}{(2\pi)^d}\partial_t
R_k\biggl(-\underbrace{\functder{J}{W_k}}_{-\phi}
	\underbrace{\functder{J}{W_k}}_{-\phi}-
	i\functderder{J}{J}{W_k}\biggr)\\
	=&-\Delta S_k-\frac i2\int \frac{d^d q}{(2\pi)^d}
	\partial_tR_k\functderder{J}{J}{W_k}\,.
\end{align*}
With this relation the variation of the effective action Eq.~\eqref{eq:GammaFlow1} 
takes the form
\begin{align*}
\partial_t\Gamma_k[\phi]=\frac i2\int \frac{d^d q}{(2\pi)^d} (\partial _t R_k)  \functderder{J}{J}{W_k}\,.
\end{align*}
The second functional derivative of $W$ with respect to the source $J$ can be written in terms 
of the effective action:
\begin{align}
\functderder{\phi}{\phi}{\Gamma_k}&=-\functder{\phi}{J}-R_k\nonumber  \\
\Rightarrow \functder{\phi}{J}&=-\left(\functderder{J}{J}{W_k}\right)^{-1}
=-\left(\functderder{\phi}{\phi}{\Gamma_k}+R_k\right)\,.
\end{align}
Making use of 
\begin{align*}
	\delta(q-q')=\functder{\phi(q')}{\phi(q)}
		=&-\functder{\phi}{}\functder{J}{W_k}
		=-\int\frac{d^d q}{(2\pi)^d}\functderder{J}{J}{W_k}\functder{\phi}{J}\,,
\end{align*}
we  obtain the Wetterich equation in Minkowski-Space:
\begin{equation}
\partial_t\Gamma_k[\phi]=\frac i2\,{\rm Tr}\,\biggl[(\partial_t R_k)
	\Bigl(\functderder{\phi}{\phi}{\Gamma_k}+R_k\Bigr)^{-1}\biggr]\,.\label{Wett:minkowski}
\end{equation}

\section{Derivation of the flow equations at finite temperature}
\label{sec:FiniteTemp}
In order to preserve supersymmetry in the RG flow for vanishing temperature, 
we must choose a regulator function which regularizes the theory in the time-like direction 
and the space-like directions in the same way. In order to make apparent how soft SUSY-breaking 
due to finite temperature emerges, we use the same regulator for our finite-temperature 
and zero-temperature studies. It is given by:
\[
r_2=\left(\frac{k}{|p|}-1\right)\theta\left(\frac{p^2}{k^2}-1\right),\quad
r_1=0\,.\]
In the LPA, the finite-temperature flow equations can be obtained straightforwardly 
from the zero-temperature flow equations by replacing $p_0$ by the Matsubara modes $\nu_n$ and $\omega_n$ 
of fermion and boson fields, respectively, and replacing the integration over $p_0$ by a summation over the 
Matsubara modes. The contribution of the bosons to the RG flow of our Wess-Zumino model then reads:
\begin{align*}
	\partial_kW_k'=W_k'''T
	\sum_{n=-\infty}^\infty\int
	\frac{d^2p_s}{8\pi^2}
	\frac{(W''^2_k-k^2)\theta(k^2-p_s^2-\omega_n^2)}
	{(k^2+W''^2_k)^2\sqrt{p_s^2+\omega_n^2}}\,,
\end{align*}
where $p_s$ denotes the momenta in space-like directions. Along the
lines of, e.\,g.,~\cite{Gies:1999vb}, we use Poisson's sum formula,
\[
	\sum_{n=-\infty}^\infty f(n)=\sum_{\ell=-\infty}^{\infty}\int_{-\infty}^\infty dqf(q)\exp(-2\pi i\ell q)\,,
\]
in order to obtain
\begin{align*}
	\partial_kW_k'=&W_k'''T\sum_{\ell=-\infty}^\infty\int
	_{-\infty}^{\infty} dq\int \frac{d^2p_s}{8\pi^2}\times\\
	&\frac{(W''^2_k-k^2)\theta(k^2-p_s^2-(2\pi
	qT)^2)}{(k^2+W''^2_k)^2\sqrt{p_s^2+(2\pi qT)^2}}{\rm e}^{-2\pi i q\ell}\,.
\end{align*}
For the computation of the three-dimensional integral, we first substitute  $q'=2\pi Tq$
and introduce spherical coordinates
\[p_s^1=r\cos\vartheta\sin\varphi,\;p_s^2=r\sin\vartheta\sin\varphi,\;q'=r\cos\vartheta.\]
Performing the angular integrations yields 
\begin{align*}
	\partial_kW_k'=W_k'''T\sum_{\ell=-\infty}^\infty\int_0^k
	\frac{dr}{8\pi^2 T} \frac{(W''^2_k-k^2)}{(k^2+W''^2_k)^2}
	\frac{2T\sin(\frac{\ell r}{T})}{\ell}\,.
\end{align*}
Finally the integration over $r$ leads to
\begin{align*}
	\partial_kW_k'=\frac{(W''^2_k-k^2)W_k'''}{8\pi^2
	(k^2+W''^2_k)^2}\,2T^2\sum_{\ell=-\infty}^\infty \frac{1-\cos(\frac {\ell
	k}T)}{\ell^2}\,.
\end{align*}
Since we made use of Poisson's resummation formula to rewrite the sum 
over the thermal modes, we are able to split the flow equation into 
a zero-temperature and a finite-temperature contribution:
\begin{align*}
	\partial_kW_k'=\frac{(W''^2_k-k^2)W_k'''}{8\pi^2 (k^2+W''^2_k)^2}
	\biggl(k^2+\underbrace{4T^2\sum_{\ell=1}^\infty
	\frac{1-\cos(\frac{k\ell}T)}{\ell^2}}_{g_{\rm bos}(T)}\biggr)\,.
\end{align*}
The contribution of the fermions to the RG flow of our model can be obtained along the lines of our derivation of the 
bosonic contribution and reads:
\begin{align*}
	\partial_kW_k'=\frac{(W''^2_k-k^2)W_k'''}{8\pi^2(k^2+W''^2_k)^2}
	\biggl(\!k^2+\underbrace{4T^2\sum_{\ell=1}^\infty(-)^\ell
	\frac{1-\cos(\frac{k\ell}T)}{\ell^2}}_{g_{\rm ferm}(T)}\biggr)\,.
\end{align*}
Introducing the dimensionless temperature $\tilde{T}=T/k$, the functions $g_{\rm bos}(T)$ and $g_{\rm ferm}(T)$ 
can be written in terms of Polylogarithms: 
\begin{align*}
	g_{\rm bos}(\T)=&\frac23 k^2\T^2\left[\pi^2-3\Li_2 ({\rm e}^{-{\rm i}/\tilde{T}}) -3\Li_2({\rm e}^{{\rm i}/\tilde{T}})
	\right]\,,\\
	g_{\rm ferm}(\T)=&-\frac26 k^2\T^2\left[\pi^2+6\Li_2(-{\rm e}^{-{\rm i}/\tilde{T}})+6\Li_2(-{\rm e}^{{\rm i}/\tilde{T}})
	\right]\,.
\end{align*}
Using the identity~\cite{Lewin:1981} 
\begin{align*}\Li_2\left(-z\right)+\Li_2\left(-\frac1z\right)&=2\Li_2(-1)-\frac12\ln^2(z)\\&=-\frac{\pi^2}{6}-\frac12\ln^2(z)\,,
\end{align*}
the function $g_{\rm bos}(T)$ simplifies further to
\begin{align}
	g_{\rm bos}(\T)
	&=\T^2\left[\pi^2+\ln^2\big(-
\exp(i/\T)\big)
 	\right]\nonumber\\
	&= \pi T\left(\pi T -(2s_B+1)^2\pi T+(2s_B+1)2k\right)-k^2\,,
\end{align}
where we have used that
\begin{align*}
	\ln\left(\exp(i/\T+i\pi)\right)=\frac i\T-i\pi\,(2s_B+1),\\
s_B\leq
	\frac{1}{2\pi\T}\leq s_B+1\;\Rightarrow\;
	s_B=\left\lfloor\frac{k}{2\pi T}\right\rfloor\,.
\end{align*}
Similarly, exploiting the relation
\begin{align*}
&	\ln\left(\exp(i/\T)\right)=\frac i\T-2i\pi \,s_F,\\&
s_F-\frac12\leq \frac{1}{2\pi\T}\leq s_F+\frac12 \;\Rightarrow\;
	s_F=\left\lfloor\frac{k}{2\pi T}+\frac12\right\rfloor\,,
\end{align*}
we end up with the result
\begin{align*}
	g_{\rm ferm}(T)
	=-k^2\left(1- s_F\frac{2\pi T}{k}\right)^2
\end{align*}
for the fermions. As expected, the functions $g_{\rm bos}(T)$ and $g_{\rm ferm}(T)$ exhibit the  same
behavior as the threshold functions discussed in Ref.~\cite{Litim:2001up}. 


\end{document}